\documentclass[final]{aipproc}
\layoutstyle{6x9}
\usepackage{graphicx}
\usepackage{epsf}
\usepackage{wrapfig}
\usepackage{sublabel}
\usepackage{epsfig}
\usepackage{amsmath}

\newcommand{\beq}{\begin{eqnarray}}
\newcommand{\eeq}{\end{eqnarray}}

\newcommand{\be}{\begin{equation}}
\newcommand{\ee}{\end{equation}}

\def\b0{{\mbox{\boldmath$0$}}}

\def\b0{{\mbox{\boldmath$0$}}}

\def \b #1{ {\bf #1}}

\def \b #1{ {\bf #1}}

 \newcommand\beqn{\begin{eqnarray}}
 \newcommand\eeqn{\end{eqnarray}}

\def\beqy{\begin{eqnarray}}
\def\eeqy{\end{eqnarray}}

\def\Re{\,\mbox{Re}\,}

\def\sqelha{\sigma_{qel}^{hA}}
\def\sthn{\sigma_{tot}^{hN}}

\def\stota{\sigma_{tot}^{hA}}
\def\sela{\sigma_{el}^{hA}}

\def\sinha{\sigma_{in}^{hA}}

\def\eeq{\end{equation}}

\def\beqy{\begin{eqnarray}}
\def\eeqy{\end{eqnarray}}

\newcommand{\ber}{\begin{displaymath}}
\newcommand{\eer}{\end{displaymath}}
\newcommand{\bey}{\begin{eqnarray}}
\newcommand{\eey}{\end{eqnarray}}

\def\beqy{\begin{eqnarray}}
\def\eeqy{\end{eqnarray}}

\begin{document}
\vskip -2mm
\date{\today}
\vskip -2mm
\title{Nucleon-Nucleon Short-Range Correlations and Gribov Inelastic Shadowing
in High Energy Hadron-Nucleus Collisions \footnote{Talk given at DIFFRACTION 2012,
{\it International Workshop on Diffraction in High-Energy Physics},
Lanzarote,
Canary Islands (Spain), September 10-15, 2012.}
}
\author{Claudio Ciofi degli Atti}{
  address={
    Istituto Nazionale di Fisica Nucleare, Sezione di Perugia,
    Via A. Pascoli, I-06123 Perugia, Italy}=thanks{Work in collaboration with UTFSM, Val  paraiso,
    CHILE}}
\begin{abstract}
Different types of high-energy hadron-nucleus cross sections are discussed emphasizing
the role played by
Nucleon-Nucleon (NN) Short-Range Correlations (SRC) and Gribov Inelastic Shadowing (IS).
\end{abstract}
\maketitle

\begin{figure}
\centerline{\includegraphics[width=6.0cm,height=5.0cm]{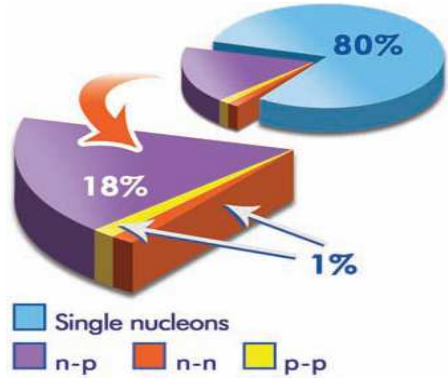}
 \vspace{-0.5cm}\caption{The percentage of NN SRC found experimentally
 in $^{12}C$ nucleus (after Ref. \cite{Subedi}).}}
  \label{Fig1}
\end{figure}

\begin{figure}
\vspace{-0.5cm}
\centerline{\includegraphics[width=8.0cm,height=6.8cm]{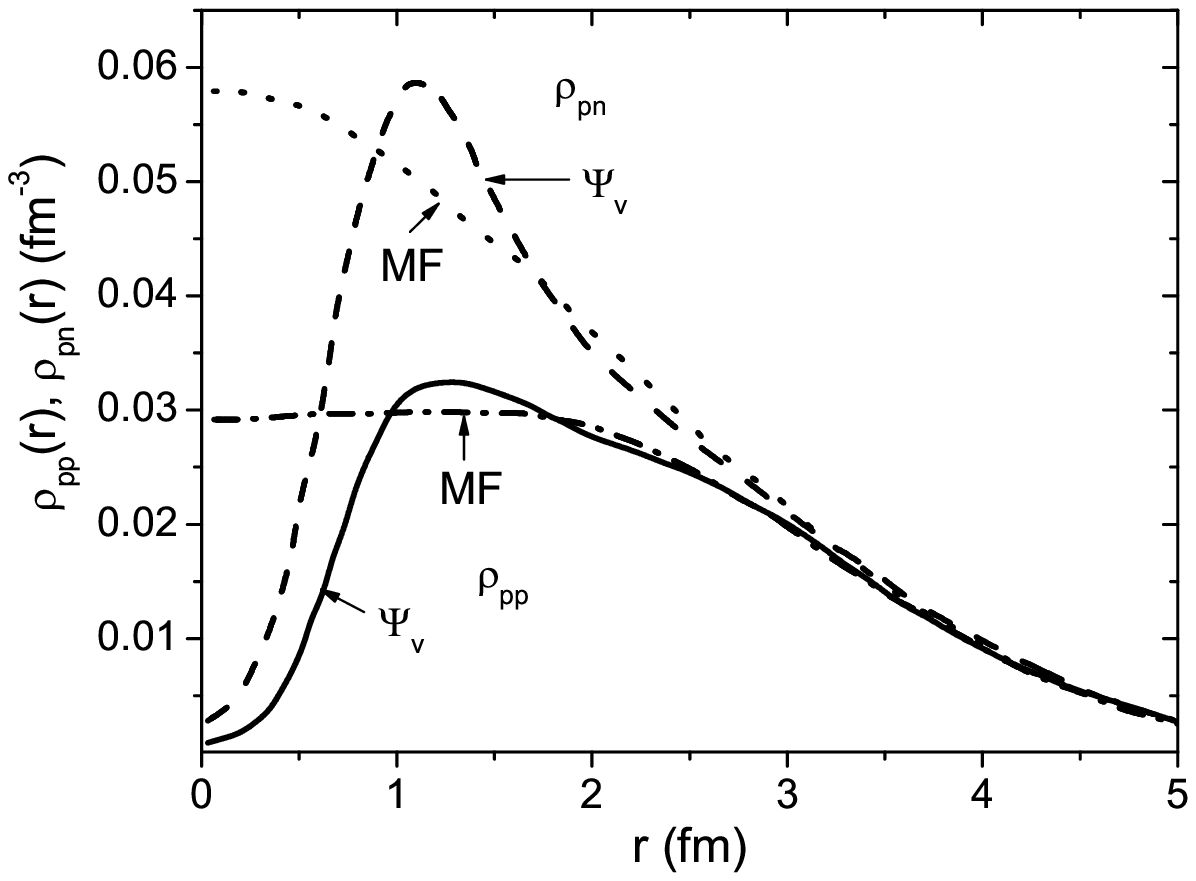}
\includegraphics[width=6.5cm,height=6.3cm]{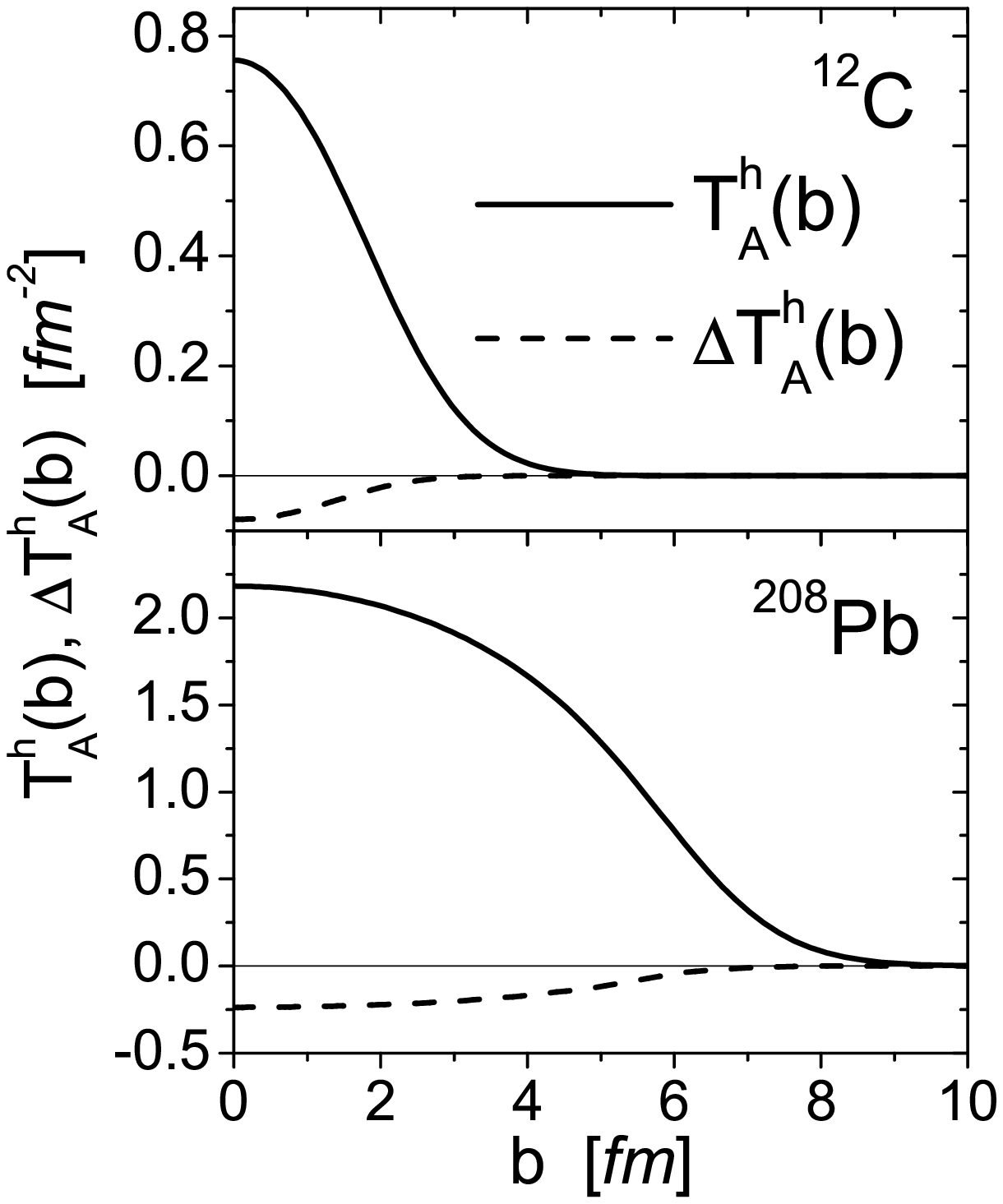}
\caption{({\it Left}) The proton-neutron $(pn)$ and proton-proton
$(pp)$ two-body densities $\rho_2\equiv\rho_{NN}$ in $^{16}O$ calculated \cite{panda}
within a mean field model (MF) and by solving the many-body problem
with a realistic NN interaction ($\Psi_v$) (after Ref. \cite{panda}). ({\it Right})
The thickness function $T_A^h(b)$ and the correlation
contribution, $\Delta T_A^h(b)$,   in $p-^{12}C$ and $p-^{208}Pb$
collisions  at HERA-B energies. The total thickness function is
given by ${\widetilde T}_A^h= T_A^h- \Delta T_A^h$ (after Ref.
\cite{nashboris}).}}
  \label{Fig2}
  \vspace{-0.5cm}
\end{figure}

 A large number of theoretical  approaches aimed at describing
hadron-nucleus  collisions at   multi $GeV$ energies,
are based upon Glauber multiple scattering theory \cite{glauber} within the
independent-particle model description of the nucleus. The latter
approximation seems to be out of date, for  nuclear constituents spend part of
their time in strongly correlated configurations \cite{panda}, as quantitatively demonstrated
by recent experimental data  \cite{Subedi} (see Fig. \ref{Fig1}). Besides NN SRC,
 intermediate hadron-hadron
inelastic scattering (Gribov  IS \cite{gribov}),
lacking in the Glauber approach,  should be taken into account.  As a matter of fact,
the importance of the effects of both SRC and Gribov IS  in different
 high energy scattering processes have
been
studied in a series of recent papers \cite{totalnA}-\cite{ennecoll}, finding that the two effects act frequently in the opposite directions.
The nuclear quantity entering most Glauber-like calculations is
  the modulus squared of the nuclear wave function $|\psi_0|^2$,
   whose exact
expansion
 \cite{glauber}
 is usually approximated by the lowest order, fully
uncorrelated term, {\it viz}
 \beqn
 \left|\,\psi_0({\bf
r}_1,...,{\bf r}_A)\,\right|^2=\prod_{j=1}^A\,\rho_1({\bf r}_j)
\,+\,\sum_{i<j}\,\Delta({\bf r}_i,{\bf
r}_j)\hspace{-0.1cm}\prod_{k\neq i,j}\rho_1({\bf
r}_k)\,+.....\simeq \prod_{j=1}^A\,\rho_1({\bf r}_j).
\label{psiquadro}
 \eeqn
\noindent Here the \textit{two-body contraction} ${\Delta({\bf r}_i,{\bf
r}_j)}\,=\,\rho_2({\bf r}_i, {\bf r}_j)\,-\,\rho_{1}({\bf
r}_i)\,\rho_{1}({\bf r}_j)$, contains the effect of SRC,
leading to a hole in  the two-body density  $\rho_2$ at short
inter-nucleon separations $r=|{\bf r}_i -{\bf r}_j|$ (see Fig. \ref{Fig2}).
SRC  generate an
additional contribution to the  nuclear thickness function as
follows
  (${\bf r}_i=\{{\bf s}_i,z_i\}$) \cite{totalnA}
\be
  \Delta T_A^h(b)=
  \frac{1}{\sthn}
  \int d^2{\bf s}_1\,d^2{\bf s}_2\,
  \Delta^\perp_A({\bf s}_1,{\bf s}_2)
  \Re \Gamma^{pN}({\bf b}-{\bf s}_1)\,
  \Re \Gamma^{pN}({\bf b}-{\bf s}_2),
  \label{55}
  \eeq
\begin{figure}
  \centerline{
\includegraphics[width=10.8cm,height=10.5cm]{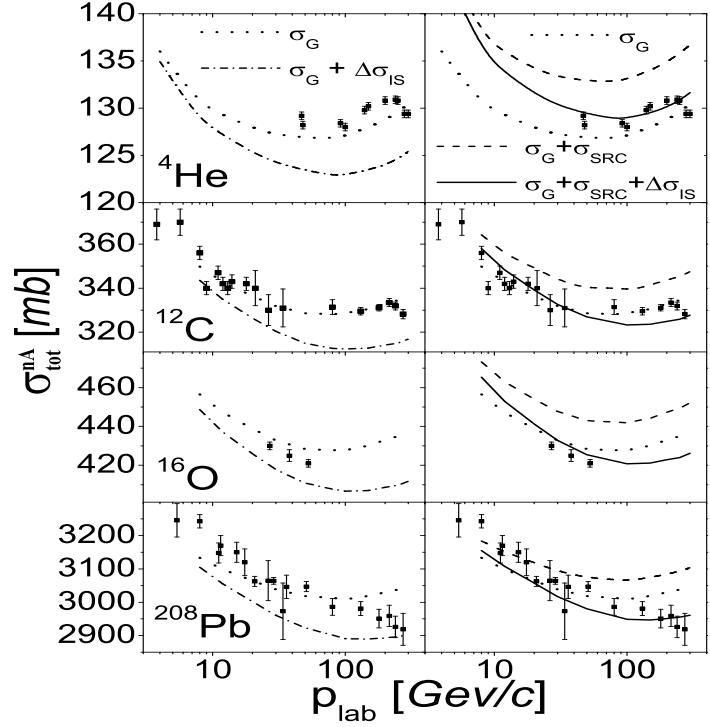}
\caption{The total neutron-nucleus cross section $\sigma_{tot}^{nA}$ {\it vs} $p_{lab}$. \textit{Left panel}:
  Glauber single density approximation ($\sigma_G$; \textit{dots})
  and Glauber plus Gribov inelastic shadowing
  ($\sigma_G$ + $\Delta\sigma_{IS}$; \textit{dot-dash}).
  \textit{Right panel}:
  Glauber ($\sigma_G$; \textit{dots});
  Glauber   plus SRC ($\sigma_G$ + $\sigma_{SRC}$; \textit{dashes});
  Glauber   plus SRC
  plus Gribov inelastic shadowing  ($\sigma_G$ + $\sigma_{SRC}$ + $\Delta\sigma_{IS}$;
    \textit{full})
  (after Ref. \cite{totalnA}).}}
\label{Fig3}
\end{figure}
\noindent where $\Delta^\perp_A({\bf s}_1,{\bf s}_2)$
  is the transverse  contraction and the total thickness function
is ${\widetilde T}_A^h= T_A^h- \Delta T_A^h$. The thickness
functions of $^{12}C$ and $^{208}Pb$ at HERA-B energies are given
in Fig. \ref{Fig2}.
In Ref. \cite{totalnA}, it has been shown that
SRC increase  the total neutron-nucleus cross section at high
energies, making the nucleus more opaque,
 unlike  Gribov IS corrections (considered in Ref. \cite{totalnA} in lowest order)
  which increase nuclear transparency (see Fig. \ref{Fig3}).
 An exhaustive calculation  \cite{nashboris} of the total, $\stota$, elastic,
$\sela$, quasi-elastic, $\sqelha$, inelastic, $\sinha$, and
diffractive dissociation cross sections, which
include both SRC and Gribov
IS,  summed to all orders by the dipole approach
\cite{KLZ,boris1,boris2},  confirms the opposite roles played by SRC
and IS (Table 1).
\begin{table}[!htp]
 \begin{tabular*}{0.78\textwidth}{@{\extracolsep{\fill}}c| c c c c }\hline\hline
$^{208}Pb$& Glauber & Glauber & q-2q model & 3q model \\
& & +SRC & +SRC & +SRC \\
\hline $\sigma_{tot}^{NA}\,\,\,  [mb]$ &3850.63 & 3885.77 & 3833.26 & 3839.26 \\
\hline $\sigma_{el}^{NA}\,\,\,[mb]$  & 1664.76 & 1690.48 & 1655.70 & 1660.67\\
\hline $\sigma_{qe}^{NA}\,\, \, [mb]$  &  120.92 &  112.65 & 113.37 & 113.88\\
\hline\hline
\end{tabular*}
 \caption{Various $p-^{208}Pb$ cross sections at LHC energies (after Ref. \cite{nashboris}).}
\end{table}
 The  thickness
function due to SRC and Gribov IS, which depends also  upon the dipole transverse
dimensions ${\bf r}_T$, the
light cone variable $\alpha$,
and the dipole-nucleon cross section and profile
function  has the
form
 \begin{eqnarray}
  &&\hspace{-0.5cm}\Delta T_A^{dip}( b,{\bf r}_T,\alpha)=\nonumber\\
  &&\hspace{-0.5cm}=\frac{1}{\sigma_{dip}(r_T)}
  \int d^2{\bf s}_1\,d^2{\bf s}_2\,
  \Delta^\perp_A({\bf s}_1,{\bf s}_2)
  \Re \Gamma^{{\bar q} q,N}({\bf b}-{\bf s}_1,{\bf r}_T,\alpha)
  \Re \Gamma^{{\bar q} q,N}({\bf b}-{\bf s}_2,{\bf r}_T,\alpha).
  \label{500}
  \end{eqnarray}
SRC and Gribov IS affect also the number of inelastic
collisions $N_{coll}=A\,\sigma_{in}^{hN}/\sigma_{in}^{hA}$  which is the normalization
factor  used to obtain the nucleus to nucleon ratio of the cross
section of a hard reaction. The results of calculations \cite{nashboris,ennecoll}
 based upon
realistic
  one- and two-body
densities and correlation functions  \cite{ACMprl}, are shown in Table 2.
The behavior of $N_{coll}$ is entirely governed by the
non-diffractive $\sigma_{in}^{NA}$ which is
decreased by SRC and increased by Gribov IS.
\begin{table}[!h]
\begin{tabular*} {0.98 \textwidth}{@{\extracolsep{\fill}}c| c
c c c c c c}\hline\hline & & & GLAUBER & & & \\ \hline &
$\sigma_{in}^{NN}\:[mb]$ & $\sigma_{tot}^{NA}\:[mb]$ &
$\sigma_{el}^{NA}\:[mb]$ & $\sigma_{qel}^{NA}
\:[mb]$ & $\sigma_{in}^{NA}\:[mb]$& $N_{coll}$\\
\hline RHIC &42.10  &3297.56 &1368.36 &66.06 &1863.14 &4.70 \\
\hline LHC  &68.30  & 3850.63 &1664.76 &120.92 &2064.95 &6.88\\
\hline \hline & & & GLAUBER+SRC & & & \\ \hline &
$\sigma_{in}^{NN}\:[mb]$ & $\sigma_{tot}^{NA}\:[mb]$ &
$\sigma_{el}^{NA}\:[mb]$ &
$\sigma_{qel}^{NA}\:[mb]$ & $\sigma_{in}^{NA}\:[mb]$  & $N_{coll}$\\
\hline RHIC &42.10  &3337.57 &1398.08 &58.47 &1881.02 &4.65\\
\hline LHC  &68.30  & 3885.77& 1690.48&112.65 & 2082.64&6.82\\
\hline \hline & & & GLAUBER+SRC+GRIBOV($q-2q$)& & & \\ \hline
& $\sigma_{in}^{NN}\:[mb]$ & $\sigma_{tot}^{NA}\:[mb]$ & $\sigma_{el}^{NA}\:[mb]$ & $\sigma_{qel}^{NA}\:[mb]$ & $\sigma_{in}^{NA}\:[mb]$  & $N_{coll}$\\
\hline RHIC &42.10  &3228.11 &1314.04 &71.99 & 1842.08 &4.75 \\
\hline LHC  &68.30  &3833.26 &1655.70 &113.37 & 2064.19 &6.88 \\
\hline \hline
 \end{tabular*}
 \caption{Number of inelastic collisions $N_{coll}$  in $p-^{208}Pb$ scattering at
  RHIC and LHC energies (after Ref. \cite{ennecoll}).}
\end{table}

\section{Acknowledgment}
I am indebted  to Boris Kopeliovich, Irina
Potashnikova and Ivan Schmidt,
Universidad Federico Santa Mar\'ia, Valpara\'iso, Chile, and to Massimiliano Alvioli
 and Chiara
Benedetta Mezzetti, INFN, Perugia, Italy, for a fruitful collaboration.

\bibliographystyle{aipproc}

\end{document}